\documentclass[prd,preprintnumbers,onecolumn,superscriptaddress,nofootinbib,showpacs,notitlepage]{revtex4-1}
\usepackage{graphicx,graphics,epsf,epsfig,epstopdf,subfigure,color,psfrag}
\usepackage{hyperref}
\usepackage{pstricks}
\usepackage{amssymb,amsmath,bm,bbm}
\usepackage{afterpage}
\usepackage{longtable}
\usepackage{latexsym, mathrsfs}
\usepackage{url}
\usepackage{paralist}
\usepackage{bbold}
\newcommand{\be}{\begin{equation}}
\newcommand{\ee}{\end{equation}}
\newcommand{\bea}{\begin{eqnarray}}
\newcommand{\eea}{\end{eqnarray}}
\newcommand{\bec}{\begin{center}}
\newcommand{\eec}{\end{center}}
\newcommand{\nn}{\nonumber}
\newcommand{\dd}{\displaystyle}

\begin{document}

\preprint{BARI-TH/2017-711} 
\medskip

\title{Quarkonium dissociation  in a far-from-equilibrium holographic setup \\}
\author{L.~Bellantuono}
\affiliation{Dipartimento Interateneo di Fisica ``M. Merlin", Universit\`a  e Politecnico di Bari, Italy}
\affiliation{Istituto Nazionale di Fisica Nucleare, Sezione di Bari, Italy}
\author{P.~Colangelo}
\affiliation{Istituto Nazionale di Fisica Nucleare, Sezione di Bari, Italy}
\author{F.~De~Fazio}
\affiliation{Istituto Nazionale di Fisica Nucleare, Sezione di Bari, Italy}
\author{F.~Giannuzzi}
\affiliation{Istituto Nazionale di Fisica Nucleare, Sezione di Bari, Italy}
\author{S.~Nicotri}
\affiliation{Istituto Nazionale di Fisica Nucleare, Sezione di Bari, Italy}

\begin{abstract}
The real-time  dissociation of the heavy quarkonium in a  strongly coupled boost-invariant non-Abelian plasma relaxing towards equilibrium is analyzed  in a  holographic framework. The effects driving the plasma out of equilibrium are described by  boundary quenching,  impulsive variations of the boundary metric. Quarkonium is represented by a classical string with endpoints kept close to the boundary. The evolution of the string profile is computed in the time-dependent geometry, and the dissociation time is evaluated for different  configurations with respect to the  direction of the plasma expansion. Dissociation occurs quickly for the quarkonium placed in the transverse plane.
\end{abstract}

\maketitle

The discovery that the  QCD matter produced in ultrarelativistic heavy ion collisions behaves as a  high-temperature, nearly frictionless, strongly coupled fluid 
raises new questions, both to  experiment and theory  \cite{Braun-Munzinger:2015hba}.  
On the one hand, through the study of new production modes and observables it is important to identify the smallest size of the QCD system characterized by this behavior, interpreting the hints of collective features 
observed  in high energy $p p$ and $p A$ collisions \cite{Schukraft:2017nbn}.
On the other hand, the real-time dynamics of the relaxing fluid needs to be  quantitatively described,  starting from the far-from-equilibrium state produced immediately after the collision and  relaxing  towards a hydrodynamic regime: this issue is  difficult to  face using traditional theoretical techniques, which deal either with  weakly coupled systems or, as in lattice QCD,   with systems at equilibrium in the Euclidean time.   The importance  of theoretical analyses of processes occurring in  out-of-equilibrium matter  has been recently emphasized  \cite{Romatschke:2016hle}.  Among those phenomena,  the heavy quarkonium dissociation   plays a  role in the characterization of the plasma, since it can be interpreted as a signature of deconfinement  \cite{Matsui:1986dk}.

Gauge/gravity duality   \cite{Maldacena:1997re,Witten:1998qj, Gubser:1998bc} (``holographic") methods provide tools to describe  out-of-equilibrium strongly coupled systems, mapping  the  dynamics of a  gauge system into a gravity problem.
Although the existence of a gravitational  dual of QCD has not been established,  yet,  insights can be gained by the  study of theories  sharing  various properties with QCD.  In particular, a   plasma of ${\cal N}=4$ supersymmetric Yang-Mills theory (SYM),  similarly  to the QCD plasma (QGP), is  nonconfined and strongly interacting. Adopting
 the gauge/gravity  approach the gravitational dual of ${\cal N}=4$ SYM in the large $N_c$ limit,  the ten dimensional $AdS_5 \times S^5$ space, can be considered,   with  a black-hole (BH)  to describe the   boundary theory at finite temperature
\footnote{An introduction to gauge/gravity methods can be found in  \cite{Ammon:2015wua}, and numerical  approaches are discussed  in \cite{Chesler:2013lia}. 
Applications  to the phenomenology of ultrarelativistic heavy ion  collision  are reviewed  in \cite{CasalderreySolana:2011us}. }.

In this framework, quarks can be  treated as  dual to open strings in the higher dimensional space  \cite{Karch:2002sh}.
Their in-medium  dynamics is  described  starting from the initial configuration  of a classical string in the dual space, and following  the string evolution dictated by the equations of motion resulting from the Nambu-Goto action.

 Different configurations have been considered  in a static gravitational background.  One  consists in  a string created at an initial time  at a point in space, which   evolves  into an extended string  falling down under the action  of  gravity, with the  endpoints  moving in opposite directions. In this way,   the drag force experienced by a heavy quark traveling in a strongly coupled plasma  \cite{Gubser:2006bz,Herzog:2006gh} and   the diffusion coefficient  have been evaluated \cite{CasalderreySolana:2006rq,Chernicoff:2006hi,Chesler:2014jva}.
At finite temperature the trajectories of the endpoints  approach lightlike geodesics  at early times  \cite{Chesler:2008wd,Chesler:2008uy}, and  thermalization occurs when  the BH horizon is reached. 
The largest penetration depth  $\Delta x$  reached by a quark with energy $E$  before thermalizing turns out to be an increasing function of the energy:   the  scaling law  $\Delta x \propto E^{1/3}$  has been determined  \cite{Gubser:2008as}, and  an  analytical proof that the penetration depth for a light quark is bounded by the curve $\Delta x =const \cdot  E^{1/3}$ has been provided \cite{Chesler:2008uy}.
For light quarks, the  energy loss  has been  found to be  sensitive to the initial conditions, and a peak  before  thermalization has been observed,  sometimes denoted as  explosive burst of energy 
 \cite{Chesler:2008uy}.

 For a quarkonium moving in the plasma,   the screening length 
has been evaluated by a Wilson loop computation
\cite{Liu:2006nn,Liu:2006he,Finazzo:2014rca},  and the Wilson loop in the  Vaidya geometry has also been examined \cite{Ali-Akbari:2015ooa}.  Here, we look at  the real-time evolution of 
a string stretched between two endpoints (representing the heavy quark and antiquark)  kept close to the boundary. The string falls down under  gravity and,  at finite temperature,
it can reach the BH horizon:   this is  interpreted as the quarkonium  dissociation in the  thermalized medium \cite{Lin:2006rf,Iatrakis:2015sua}.

\medskip
To describe  the  dynamics  in the out-of-equilibrium matter,   we consider the time-dependent dual geometry  obtained
 through a quench,   a distortion of the boundary metric   mimicking an impulsive effect  which drives the boundary system far from equilibrium \cite{Chesler:2008hg,Chesler:2009cy}. 
We denote the $4D$ boundary coordinates  as  $x^\mu=(x^0,x^1,x^2,x^3)$, with  $x^3=x_\parallel$     the axis in the collision direction. We study the case  where  boost invariance along this  axis is imposed, together with  translation  and $O(2)$ rotational invariance  in the     $x_\perp=\{x^1,\,x^2 \}$ plane. In terms of  the proper time $t$ and of the spatial rapidity $y$,   with $x^0=t \cosh y$, $x_\parallel=t \sinh y$,  the $4D$  line element  is written as \footnote{We denote the proper time as $t$,  using  $\tau$ for one of the string world-sheet coordinates.}:
$ds^2_4=-dt^2+dx_\perp^2+t^2 dy^2$. 
A quench is introduced as an anisotropic distortion of the boundary metric  which  leaves the spatial three-volume unchanged and maintains the other symmetries.  The line element $ds^2_4$ is modified as
\be
ds^2_4=-dt^2+e^{\gamma(t)} dx_\perp^2+t^2e^{-2\gamma(t)} dy^2 \,\,, \label{metric4D}
\ee
with  quench profile  $\gamma(t)$.
Equation~~(\ref{metric4D}) can be considered as the boundary metric of the $5D$ space where the gravity dual is defined. Using  Eddington-Finkelstein  coordinates,   with $r$ the fifth  (holographic) coordinate, the $5D$ bulk metric can be written  as
\be
ds^2_5=2 dr dt-A dt^2+ \Sigma^2 e^B dx_\perp^2+ \Sigma^2 e^{-2B}dy^2  \,\,\, , \label{metric5D}
\ee
with the boundary obtained for  $r \to \infty$.  
The metric functions
 $A$, $\Sigma$ and $B$ only depend  on $r$ and $t$, due to the imposed symmetries.  They are determined solving  $5D$ Einstein vacuum equations with negative cosmological constant, with two boundary conditions.  The first one is that the  metric  (\ref{metric5D}) reproduces  (\ref{metric4D}) for $r \to \infty$.  
Moreover,  focusing on  distortion profiles in a time interval $[t_i,\,t_f]$, the metric functions must correspond to   AdS$_5$  at $t=t_i$ \cite{Chesler:2009cy}. 

\begin{figure}[b!]
\begin{center}
\includegraphics[width = 0.45\textwidth]{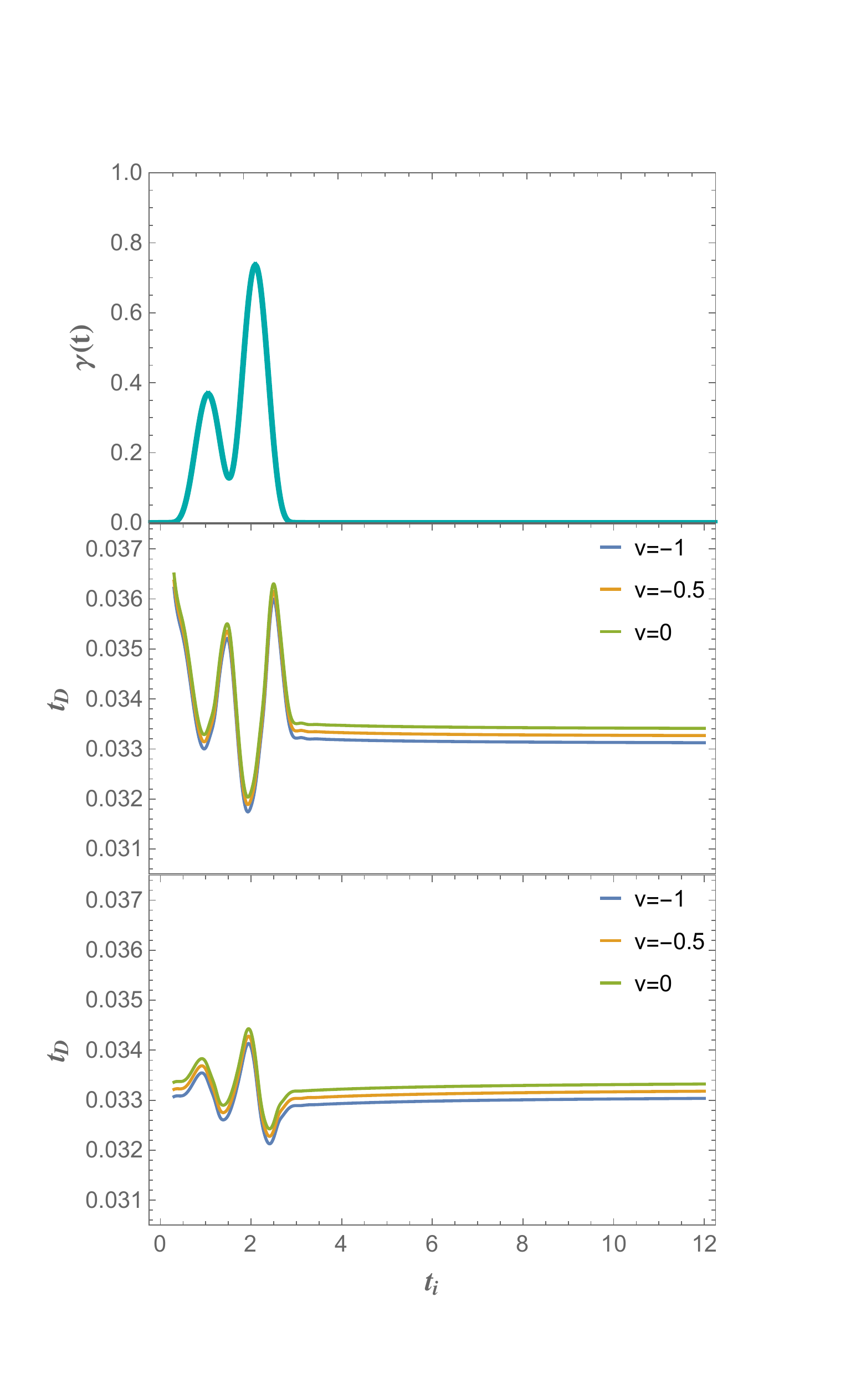}
\includegraphics[width = 0.45\textwidth]{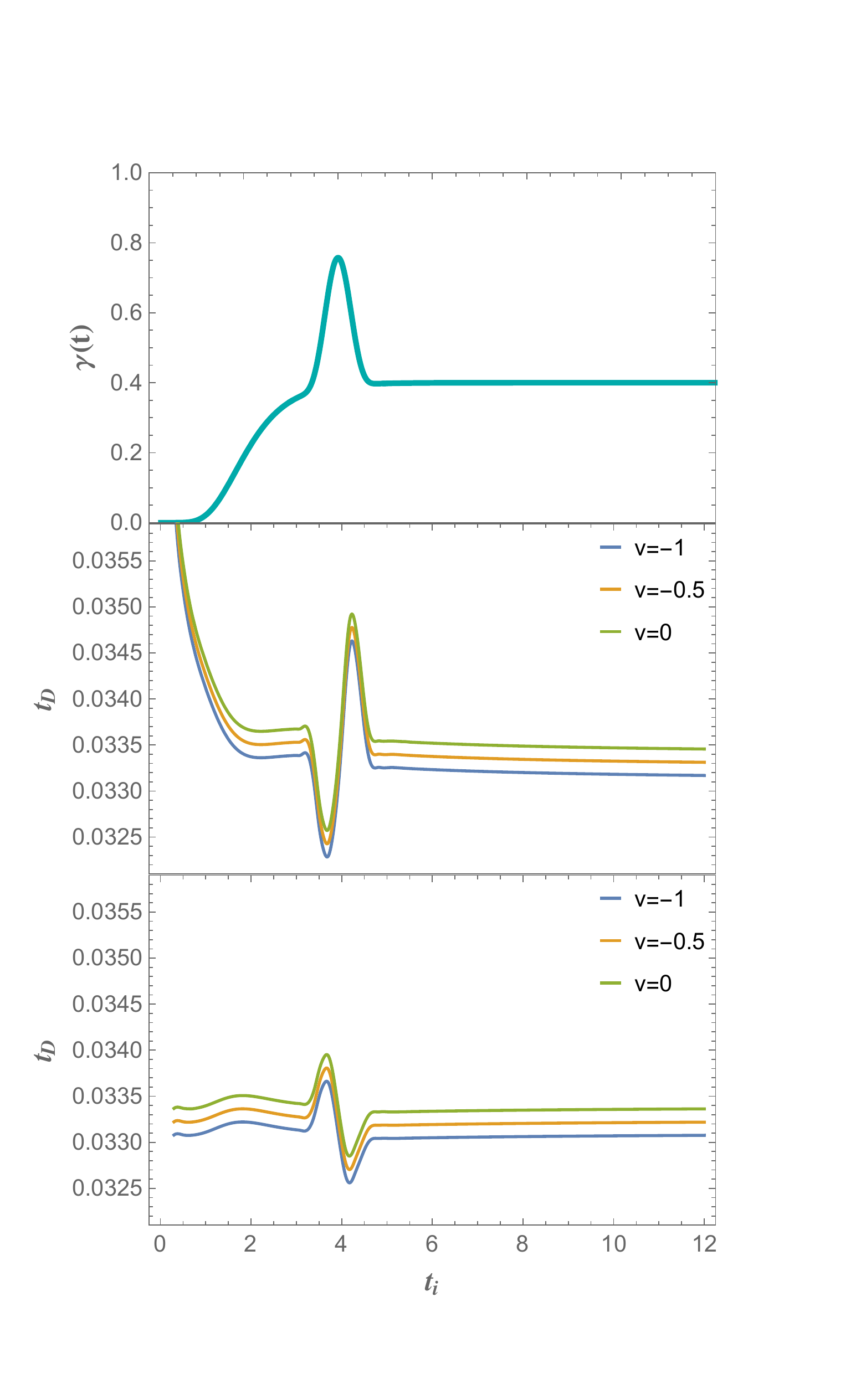}\\
\vspace*{-0.5cm}
\caption{\baselineskip 10pt  \small Quarkonium dissociation time $t_D$ versus $t_i$ in quench model $\cal A$(2) (left) and ${\cal B}$ (right).  The  quench profiles $\gamma(t)$ are drawn in the top panels. 
The middle plots refer to the $w=y$  string configuration  with   $y_L=10$,   the bottom plots to the transverse $w=x$ configuration   and   $L=10$. Results are shown for three  values of the initial velocity $v$,  indicated in the legends.  }\label{columns}
\end{center}
\end{figure}

In \cite{Bellantuono:2015hxa,Bellantuono:2016tkh} several  quench profiles have been investigated,  describing  sequences of  pulses with  different intensities and duration.
Here we focus on two cases,  one denoted as model ${\cal A}(2)$,  consisting of two overlapping  short pulses, and  one as model ${\cal B}$,  a smooth distorsion  with a    pulse superimposed \cite{Bellantuono:2015hxa}. They are obtained  choosing the profiles
\be
\gamma(t) =w_\gamma \left[ \tanh \left(\frac{t -t_0}{\eta} \right) \right]^7 \,+\sum_{j=1}^{N}\gamma_j(t,t_{0,j})  \label{profile}
\ee
with
\be
\gamma_j(t,t_{0,j}) = c_j f_j(t,t_{0,j})^6 e^{-1/f_j(t,t_{0,j})} \Theta\left(1-\frac{(t-t_{0,j})^2}{\Delta_j^2}\right)\,\, \label{profile1} 
\ee
and 
\be
f_j(t,t_{0,j})= 1- \frac{(t-t_{0,j})^2}{\Delta_j^2} \,\,\, , \label{profile2}
\ee
$\Theta$ being the Heaviside function.
The set of parameters $w_\gamma, \eta, t_0, t_{0,j},  c_j$,  $ \Delta_j$  and $N$ specifies each quench model:
for  ${\cal A}(2)$,   two overlapping short pulses,  the   parameters are 
$w_\gamma=0$,  $N=2$, 
$c_1 = 1$, $\Delta_{1,2}=1$, $t_{0,1}=\frac{5 }{4}\Delta_1$,
$c_2 = 2$,  $t_{0,2}=\frac{9}{4} \Delta_2$;  the quench ends at  $t_f^{\cal A}=3.25$.
Model $\cal B$,  a short pulse superimposed to a slow deformation,  is obtained using 
$w_\gamma=\frac{2}{5}$, $\eta=1.2$, $t_0=0.25$, $N=1$, 
$c_1 = 1$, $\Delta_1=1$, $t_{0,1}=4 \Delta_1$; the pulse ends at  $t_f^{\cal B}=5$,  while  a slow distortion continues  approaching a constant value.
In both models, with profiles $\gamma(t)$  drawn in the top panels of Fig.~\ref{columns},  the quench starts at $t_i=0.25$. 
The Einstein equations with  quenches have been solved numerically \cite{Bellantuono:2015hxa}, determining the  metric functions
 $A(r,t)$, $\Sigma(r,t)$ and $B(r,t)$  that will be used  in the analysis of the quarkonium dissociation. Computing the boundary theory stress-energy tensor for the two models,  it was found that quantities such as the  energy density follow a viscous hydrodynamics behavior immediately after the end of the spikes in the  quenches, at $t_{hydro}^{{\cal A}(2)}=3.25$ and  $t_{hydro}^{\cal B}=5$, while a pressure anisotropy $\dd \frac{p_\parallel}{p_\perp}$ persists for a longer time, up to
$t_{isotr}^{{\cal A}(2)}=6$ and  $t_{isotr}^{\cal B}=6.74$ \cite{Bellantuono:2015hxa}.

\vspace*{0.5cm}
To describe a quark-antiquark pair produced in a far-from-equilibrium medium, we consider 
 a classical string in the higher dimensional space described by the metric  (\ref{metric5D}), with  endpoints kept close to the boundary. The string dynamics  is governed by the equations of motion  from the Nambu-Goto action
\be
S_{NG}=-T_f \int d\tau d\sigma \sqrt{-g}=\int d\tau d\sigma {\cal L}_{NG}
\label{SNG}
\ee
with $T_f=\displaystyle{\frac{1}{2\pi \alpha^\prime}}$, $\alpha^\prime=\frac{L^2}{\sqrt{\lambda}}$,  $L$ the AdS$_5$ radius and $\lambda$ the `t Hooft coupling. $g$ is the determinant of the induced world-sheet metric, and $(\tau,\,\sigma)$  the world-sheet coordinates. The string profile is described by the embedding functions $X^M(\tau,\,\sigma)$, so that
\be
-g=\left[G_{MN}{\dot X}^M X^{\prime N} \right]^2- ({\dot X})^2(X^\prime)^2 \,\,,
\label{gamma}
\ee
in terms of  ${\dot X}^M=\partial_\tau X^M$ and $X^{\prime M}=\partial_\sigma X^M$, with $G_{MN}$ the metric in (\ref{metric5D}).
For strings living in a three-dimensional slice of the dual space described by the coordinates $(t,\,w,\,r)$,  we focus on two possibilities.
 The first one is  $w=x$, with $x=x_1$ or $x=x_2$  one of the two transverse coordinates, and the string endpoints kept fixed at distance $2 L$ close to the boundary. The second one is 
  $w=y$ along the rapidity axis, representing a quark and an antiquark  moving away  from each other  in the longitudinal direction $x_\parallel$ with rapidity $y_L$.

In the  gauge $\tau=t$ and $\sigma=w$ the string profile is described by the function $r(t,\,w)$.
The Nambu-Goto action can be written in terms of the metric functions in Eq.~(\ref{metric5D}):
\be
S_{NG}=-T_f \int dt \,dw \sqrt{-g} =-T_f \int dt \, dw \sqrt{\Sigma_w(t,r)\left(A(t,r)-2\, \partial_t r\right)+\left( \partial_w r \right)^2 }\,\,\, ,  \label{SNGnoi}
\ee
where    $\Sigma_w={\bar \Sigma}=\Sigma^2 e^{-2B}$ if $w=y$, and $\Sigma_w={\tilde \Sigma}=\Sigma^2 e^{B}$ if $w=x$.
Defining ${\cal L}_{NG}=-T_f \, \sqrt{-g}$ in (\ref{SNGnoi}),  the canonical momentum  densities of the string  can be computed,
\be
\pi_M^a=\frac{\partial {\cal L}_{NG}}{\partial (\partial_a X^M)} \,\, ,
\label{Pi}
\ee
obtaining
\begin{equation}  \label{pi-densities}
 \left(\begin{array}{cc}
 \pi_t^0 & \pi_t^1 \nonumber \\
  \pi_w^0 & \pi_w^1 \nonumber \\
   \pi_r^0 & \pi_r^1 \nonumber 
\end{array}\right)
=
- \frac{T_f}{\sqrt{-g}} \left(\begin{array}{cc}
 \Sigma_w(A-{\dot r})+r^{\prime 2} &-  r^\prime {\dot r} \nonumber \\
  \Sigma_w  r^\prime & \Sigma_w(A-2{\dot r})  \nonumber \\
  -\Sigma_w & r^\prime \nonumber
\end{array}\right) \,\,  ,
\end{equation}
and the equations of motion 
\be 
\frac{\partial {\cal L}_{NG}}{\partial X^M} - \frac{d}{d\tau} \pi_M^0 - \frac{d}{d\sigma} \pi_M^1=0 \,\,\, .
\label{ELX}
\ee

\medskip
We pause here, before considering the time-dependent geometry, since it is interesting to analyze  three simpler cases
of a  5D geometry  with  Eddington-Finkelstein coordinates as  in (\ref{metric5D}), and the dynamics of a  string with profile  described by a function $r(t)$  depending  only on the time coordinate.  The three cases are  
AdS$_5$, AdS$_5$ with a BH, and AdS$_5$ with a black-brane, e.g.  a  BH with time-dependent horizon position as in a hydrodynamic dual. The
 dissociation time turns out to be finite in our coordinate system.

In pure AdS$_5$, described by metric functions  $A(r,t)={\tilde \Sigma}(r,t)=r^2$  and ${\bar \Sigma}(r,t)=r^2 t^2$,    the time required for a string to move from an initial position to  the AdS center can be easily computed.
 $ {\cal L}_{NG}$ does not depend explicitly on time,  and the conjugate momentum $ \pi_t^0$ is a constant of motion, denoted as  $p$:   $\pi_t^0=-\displaystyle{\frac{{\tilde \Sigma}(A-{\dot r})}{\sqrt{{\tilde \Sigma}(A-2{\dot r}) }}} = p$ (we  consider  the case $\Sigma_w={\tilde \Sigma}$, setting $T_f=1$).
With the initial condition  $r_0=r(t_i)$  the value of $r$ at the initial time $t_i$, and  ${\dot r}(t_i)={\dot r}_0$,  one has
\bea
{\dot r}&=&\frac{r^4-p(p-\sqrt{p^2-r^4})}{r^2} \,\,,\\
p&=&-\frac{r_0(r_0^2-{\dot r}_0)}{\sqrt{r_0^2-2{\dot r}_0}} \,\,. \label{sol-rpuntoAdS}
\eea
Starting from  $r_0$, the position $r$ is reached after the time lapse
\be
\Delta t=t-t_i=\int_{r_0}^{r} dr \frac{1}{{\dot r}}
\ee
that can be computed analytically:
\bea
\Delta t^{AdS}&=&\frac{1}{r_0 x}
 \Big\{\frac{x}{1-v_0}-1+\sqrt{1-x^4 \, q^4} \nn \\
 &-&x\,q\Big[E\left({\rm ArcSin}(q),-1 \right)-E\left({\rm ArcSin}(q x),-1 \right)
  -F\left({\rm ArcSin}(q),-1 \right) + F\left({\rm ArcSin}(q x),-1 \right)\Big] \Big\} , \label{tAdS}
 \eea
with $x=\displaystyle\frac{r}{r_0}$, $v_0=\displaystyle\frac{{\dot r}_0}{r_0^2}$,  $q=\displaystyle\frac{(1-2v_0)^{1/4}}{(1-v_0)^{1/2}}$.
 $F$ and $E$ are  elliptic integrals of the first and  second kind, respectively.
 For ${\dot r}_0=0$ the result is:
 \be
\Delta t^{AdS}=\frac{1}{r_0 x}
\Big\{-1+\sqrt{1-x^4}+x\Big[1-E(-1)+E\left({\rm ArcSin}(x),-1 \right) 
 -F\left({\rm ArcSin}(x),-1 \right)+K\left(-1 \right)\Big] \Big\} , \label{solAdS}
 \ee
 the function $E$ with a single argument being the complete elliptic integral of the second kind, and $K$  the complete elliptic integral of the first kind.
 Equation~(\ref{solAdS}) implicitly defines the string profile $r(t)$.
The AdS center is reached in a finite time lapse, since   for   $r \to 0$
\be
t_D^{AdS}=t-t_i=\frac{1}{r_0}\left(1-\frac{\sqrt{2}\pi^{3/2}}{\Gamma \left(\frac{1}{4}\right)^2}\right)\label{tcentroAdS}\,\,.
\ee

In AdS black-hole geometry  described by the metric functions 
\be
A(r,t)=r^2\left(1-\frac{r_H^4}{r^4}\right) \,\,, \hspace*{1.cm}\,  {\tilde \Sigma}(r,t)=r^2 \,\,, \hspace*{1.cm}\,  {\bar \Sigma}(r,t)=r^2 t^2 ,\label{BH}
\ee
the time interval $\Delta t$  can also be  analytically determined for a generic ${\dot r}_0$. For ${\dot r}_0=0$ one has
\be
{\dot r}=-\frac{r_0^4-r^4+\sqrt{(r_0^4-r^4)(r_0^4-r_H^4)}}{r^2} \,\, .
\label{rpuntoBH}
\ee
Using $Y=\sqrt{1-\frac{r_H^4}{r_0^4}}$ and $P=\sqrt{1-\frac{r^4}{r_0^4}}$,
the time  to reach the position $r$ starting from $r=r_0$ is 
\be
\Delta t^{BH}=\frac{2\sqrt{2}}{3r_0(1+Y)}\Big[F_1\Big(\frac{3}{4},\frac{1}{2},1,\frac{7}{4},-1,\frac{1-Y}{1+Y}\Big)
-\frac{(1-P^2)^\frac{3}{4}}{(1+P)^\frac{3}{2}}F_1\Big(\frac{3}{4},\frac{1}{2},1,\frac{7}{4},-\frac{1-P}{1+P},\frac{1-P}{1+P}\,\frac{1-Y}{1+Y}\Big)\Big]
\,, \label{solBH}
\ee
with $F_1$  the Appell hypergeometric function of two variables. 
 Also Eq.~(\ref{solBH})  implicitly defines the string profile $r(t)$.
The time required to reach the horizon $r=r_H$ is obtained substituting $P\to Y$:
\be
 t_D^{BH}=t-t_i =\frac{2\sqrt{2}}{3r_0(1+Y)^{7/4}}
 \Big[(1+Y)^\frac{3}{4} F_1\left(\frac{3}{4},\frac{1}{2},1,\frac{7}{4},-1,\frac{1-Y}{1+Y}\right) 
-(1-Y)^\frac{3}{4} F_1\Big(\frac{3}{4},\frac{1}{2},1,\frac{7}{4},-\frac{1-Y}{1+Y},\frac{(1-Y)^2}{(1+Y)^2}\Big)\Big] . 
\ee
For   $Y \to 0$ we recover  Eq.~(\ref{tcentroAdS}) for pure AdS$_5$.

The result for the black-brane geometry with time-dependent horizon can be obtained from the previous one, considering  $r_H=r_H(t)$.
The geometry dual to viscous hydrodynamics derived in \cite{Janik:2005zt} belongs to this class of geometries.  If in the interval $\Delta t$ the change in horizon position is small, 
the string profile  can still be found  substituting  $r_H \to r_H(t)$ in  Eq.~(\ref{solBH}):
\bea
\Delta t^{HYDRO}=t-t_i &=&\frac{2\sqrt{2}}{3r_0(1+Y(t))}
\Big[F_1\Big(\frac{3}{4},\frac{1}{2},1,\frac{7}{4},-1,\frac{1-Y(t)}{1+Y(t)}\Big) \nn \\
&-&
\frac{(1-P^2)^\frac{3}{4}}{(1+P)^\frac{3}{2}}F_1\Big(\frac{3}{4},\frac{1}{2},1,\frac{7}{4},-\frac{1-P}{1+P},\frac{1-P}{1+P}\,\frac{1-Y(t)}{1+Y(t)}\Big)\Big]\,  \label{solHYDRO} \eea
with
\be
Y(t)=\sqrt{1-\frac{r_H(t)^4}{r_0^4}} \,\, .  \label{solHYDRO1}
\ee
Equation~(\ref{solHYDRO}) implicitly defines the string profile $r(t)$ which solves Eq.~(\ref{ELX}) for  $M=5$  and  AdS-BH metric functions with time-dependent $r_H(t)$:
\be r(t)^2\,{\ddot  r}(t)-6 r(t)^3 {\dot r}(t) - 
 2 r(t) (r_H(t)^4 - {\dot r}(t)^2)+2r(t)^5
+2r_H(t)^3 {\dot r}_H(t)=0\,\,. \label{eomH}
\ee
This can be shown  dividing the interval  $[t_i,\,t_D]$ of  (\ref{eomH}) in equal-size intervals of infinitesimal width, $[t_i,\,t_D]=\bigcup_{k=1}^N [t_k,\,t_{k+1}]$,  where $t_{k+1}-t_k=\epsilon$, $t_0=t_i$, $t_{N+1}=t_D$, and  $\epsilon \to 0$, $N \to \infty$.
In each time interval, we solve (\ref{eomH})  with the substitution $r_H(t)=r_H(t_k)=r_H(t_{k+1})$,
 provided that the   $r_H(t)$ changes with time  slower  than the solution $r(t)$, obtaining that Eq.~(\ref{eomH}) coincides with the corresponding one  for a  static AdS-BH geometry, 
 \be r(t)^2\,{\ddot  r}(t)-6 r(t)^3 {\dot r}(t) - 
 2 r(t) (r_{H}(t_k) - {\dot r}(t)^2)+2r(t)^5
=0\,\,. \label{eomBH}
\ee
The implicit solution of (\ref{eomBH}) is given by Eq.~(\ref{solBH}),
which solves  (\ref{eomH}) in  $[t_k,t_{k+1}]$.  Considering all  intervals and  taking  $N \to \infty$, one finds that Eq.~(\ref{solBH})  provides the solution for the black-brane characterized by the horizon $r_H(t)$. 

\begin{figure}[t!]
\begin{center}
\hspace*{0.5cm}
\includegraphics[width = 0.55\textwidth]{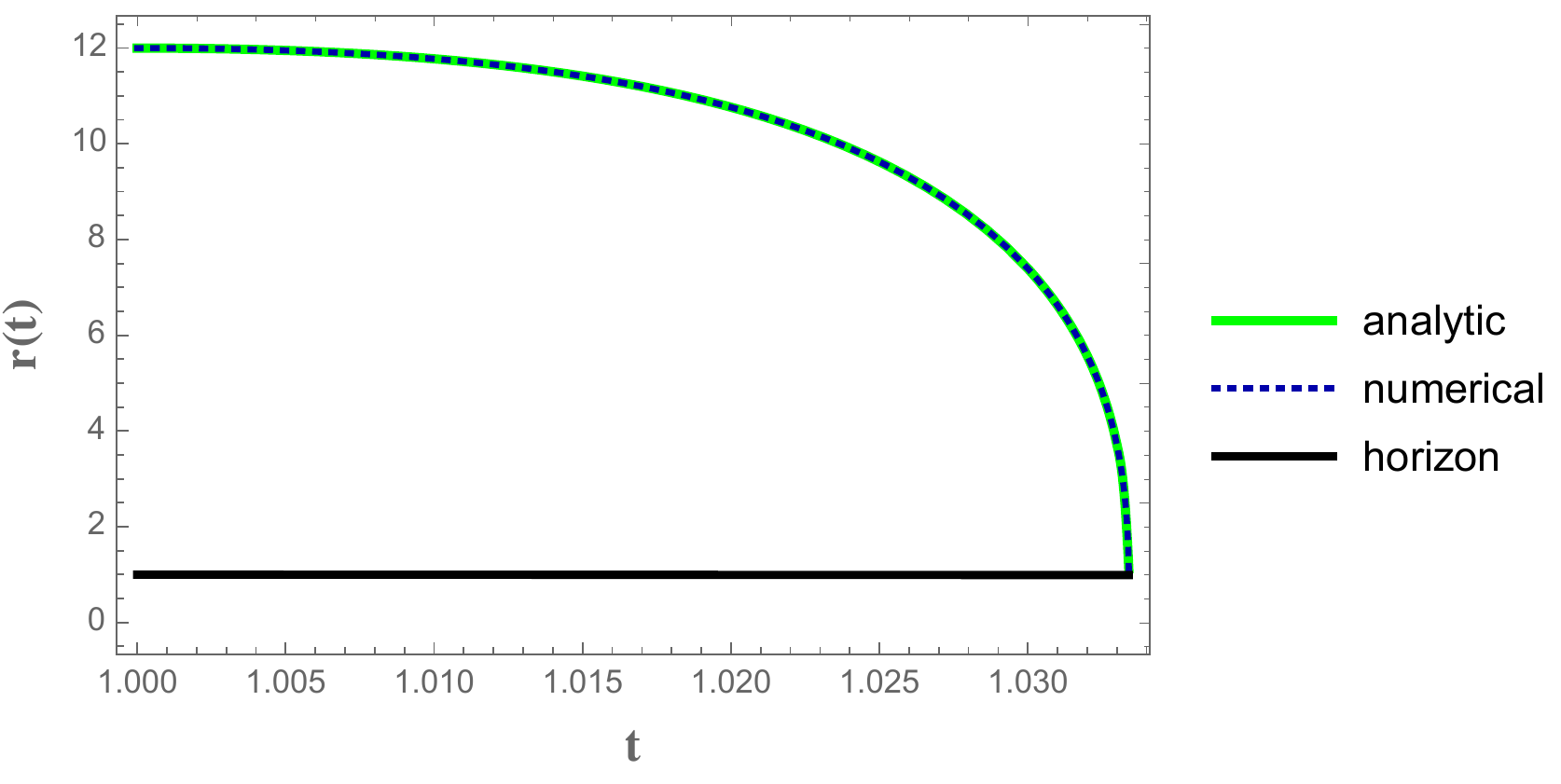}
\caption{\baselineskip 10pt \small 
String profile $r(t)$ for the AdS-BH and $t-$dependent horizon (\ref{horizon}), obtained using the expressions in (\ref{solHYDRO}),(\ref{solHYDRO1})  and the numerical solution of (\ref{eomH}) (with $r_0=12$, $t_i=1$  and $e_0=3$). }\label{check}
\end{center}
\end{figure}

  The time $t_H$ at which  the horizon is reached can be computed solving  numerically the equation $t_H-t_i=\Delta t^{HYDRO}$,  with  $r(t_H)=r_H(t_H)$. 
To  check that Eq.~(\ref{solHYDRO}) provides  the  solution of  (\ref{eomBH}), we choose, e.g., 
\be
r_H(t)=\left(\frac{e_0}{3}\right)^{1/4}\frac{1}{t^{1/3}}\,\,, \label{horizon} 
\ee
 corresponding to inviscid hydrodynamics  \cite{Janik:2005zt}, we fix $t_i=1$ and $e_0=3$
and solve numerically (\ref{eomBH}).
 The solution  coincides  with  Eq.~(\ref{solHYDRO}) as shown in Fig.~\ref{check}.
 At odds with the static cases (\ref{tAdS}) and  (\ref{solBH}), the explicit $t-$dependence in (\ref{solHYDRO}) induces a nontrivial relation between $t_D$ and  $t_i$, as we shall see in the following. 

\begin{figure}[t]
\begin{center}
\hspace*{0.5cm}
\includegraphics[width = 0.55\textwidth]{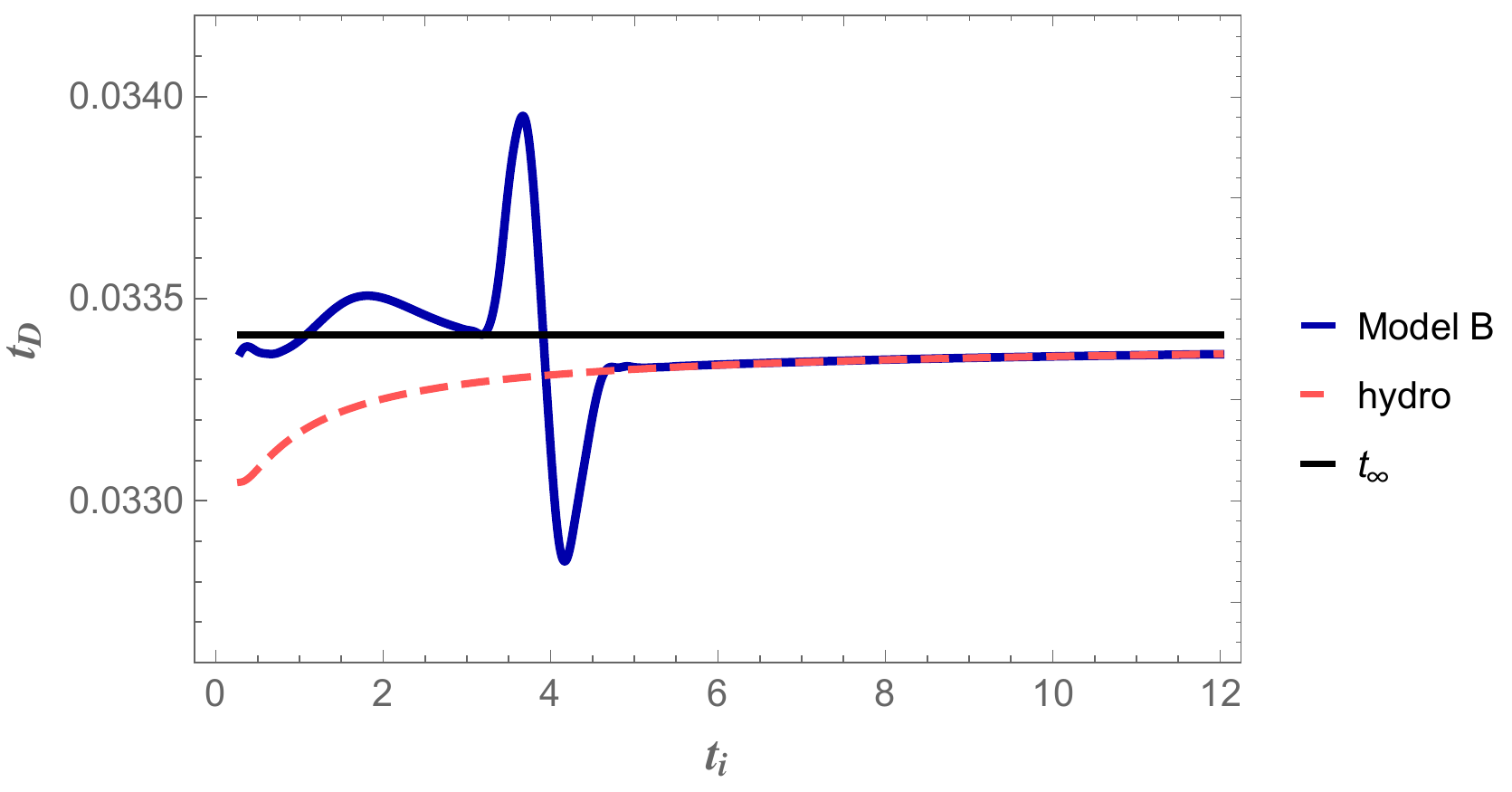}
\caption{ \baselineskip 10pt \small  Quarkonium dissociation time $t_D$ versus  $t_i$ for  the transverse  configuration  $w=x$, with separation $L=10$ and $v=0$.   The continuous line is  the result for  quench model  ${\cal B}$,
the dashed line for a geometry dual to viscous hydrodynamics, with horizon $r_H(t)$ in (\ref{Teff1}).  The  horizontal line corresponds to  the asymptotic  value in Eq.~(\ref{tas}).}\label{tDvsHydro}
\end{center}
\end{figure}

An approximate large-$t$ solution for the string profile in an AdS black-brane metric with time-dependent horizon 
is  $\dd r(t)\sim \frac{1}{t^{1/3}}$, as it can be verified 
 plugging this function in (\ref{eomH}) and observing  that  the equation   is satisfied  but  for terms  $\displaystyle {\cal O}(t^{-7/3})$.
\footnote{A similar behavior is found  for large  $t$ in the Fefferman-Graham coordinates  \cite{Lin:2006rf}.} This allows to compute  the asymptotic (large-$t$) value of the dissociation time $t_D$,
\be
t_\infty=\frac{2}{3r_0} \,_2F_1\left(1,\frac{5}{4},\frac{7}{4},-1\right)  \,\, \label{tas}
\ee
which coincides with  (\ref{tcentroAdS}).
For $r_0=12$,  an input value  in our analysis,   $t_\infty=0.0334108$ is obtained.
As it will be shown in Fig.~\ref{tDvsHydro}, this value is rapidly approached by the dissociation time  $t_D$ computed with the  metric functions $A$, $\Sigma$ and $B$ of  the geometry with quenches.
\begin{figure}[t!]
\begin{center}
\includegraphics[width = 0.47\textwidth]{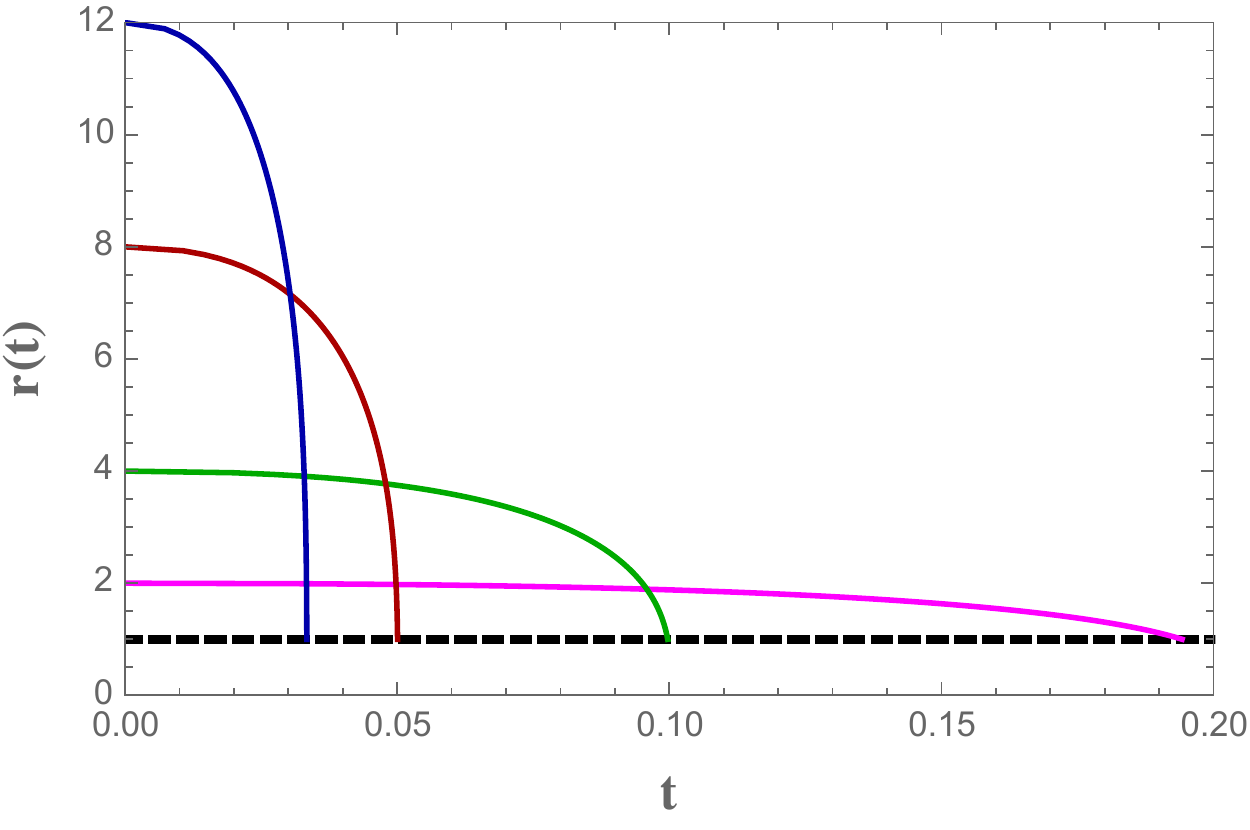}
\caption{\baselineskip 10pt \small 
String profile $r(t)$ in the AdS-BH metric (\ref{BH})  and transverse $w=x$ configuration, for  $r_0=12$ (blue line), $r_0=8$  (red), $r_0=4$ (green) and $r_0=2$ (pink). }\label{r0}
\end{center}
\end{figure}

A feature of the solution is that $t_D$ decreases when the starting point $r_0$ is closer to the boundary, as shown
 in Fig.~\ref{r0} for the AdS-BH  with $r_H=1$.  
Identifying the  position  of the string  endpoints near  the boundary   with  the quark  mass  \cite{Herzog:2006gh},  this implies that a heavier quarkoniun  dissociates faster than a lighter one.

\medskip
Let us now  consider the  metric functions  for the out-of-equilibrium system \cite{Bellantuono:2015hxa}. They are characterized by two regimes.  In the early-time regime the geometry abruptly responds to the   external quenches, a black-hole is formed with a rapidly changing horizon position  \cite{Chesler:2008hg,Chesler:2009cy}.  After the pulses are switched off,  a black-brane geometry is recovered, with  the metric functions  reconstructed in   \cite{Bellantuono:2016tkh}.  The horizon position is found with  time dependence
\be
r_{H}(t)=\frac{\pi \Lambda}{(\Lambda t)^{1/3}} \Bigg[ 1-\frac{1}{6 \pi (\Lambda t)^{2/3}}+\frac{-1+\log 2}{36 \pi^2 (\Lambda t)^{4/3} } 
+\frac{-21+2\pi^2+51 \log 2 -24 (\log 2)^2}{1944 \pi^3 (\Lambda t)^2} + {\cal O}\left( \frac{1}{(\Lambda t)^{8/3}} \right )\Bigg] \,\,\ . \label{Teff1}
\ee
  The parameters $\Lambda^{\cal B}=1.12$
and $\Lambda^{{\cal A}(2)}=1.73$  refer to the two quench model. Equation~(\ref{horizon}) is the leading term of (\ref{Teff1}). The late-time regime is dual to viscous hydrodynamics.

The  equation for the string profile   $r(t,w)$ obtained from Eqs.(\ref{SNGnoi})  and (\ref{ELX}) and  $M=5$ reads:
\be
r^{\prime \prime}-\frac{\partial_w g}{2g} r^\prime +\frac{\partial_t g}{2g}\Sigma_w-\partial_t \Sigma_w+\frac{\partial_r g}{2}=0 \,.
\label{eom} 
\ee
Choosing  $w=y$ or $w=x$, the equations can  be  solved  imposing suitable boundary and initial conditions. In our
 time-dependent metric we vary the initial time $t_i$ at which to  start considering the string evolution. At  $t_i$ the string is completely  stretched close to the boundary,  $r(t_i,\,w)=r_{max}$ for all $w$, with $r_{max}$  the maximum value of the radial coordinate used in the numerical calculation ($r_{max}=12$).
 The string endpoints  are kept  fixed at  $w_Q=-L$ and $w_{\overline Q}=L$ for the transverse configuration, and  $w_Q=-y_L$ and $w_{\overline Q}=y_L$ for the $w=y$ configuration, so that
 $r(t,w_{\overline Q})=r(t,w_{Q})=r_{max}$ for all  $t$.  We  study  the dependence on the spatial and rapidity separation  varying $L$ and $y_L$ in the range $[0.1,\,100]$.
As a further condition, we impose  $\dot r(t_i,w)=v$, with  $v=0,\,-0.5,\,-1$.

\begin{figure}[t!]
\begin{center}
\includegraphics[width = 0.45\textwidth]{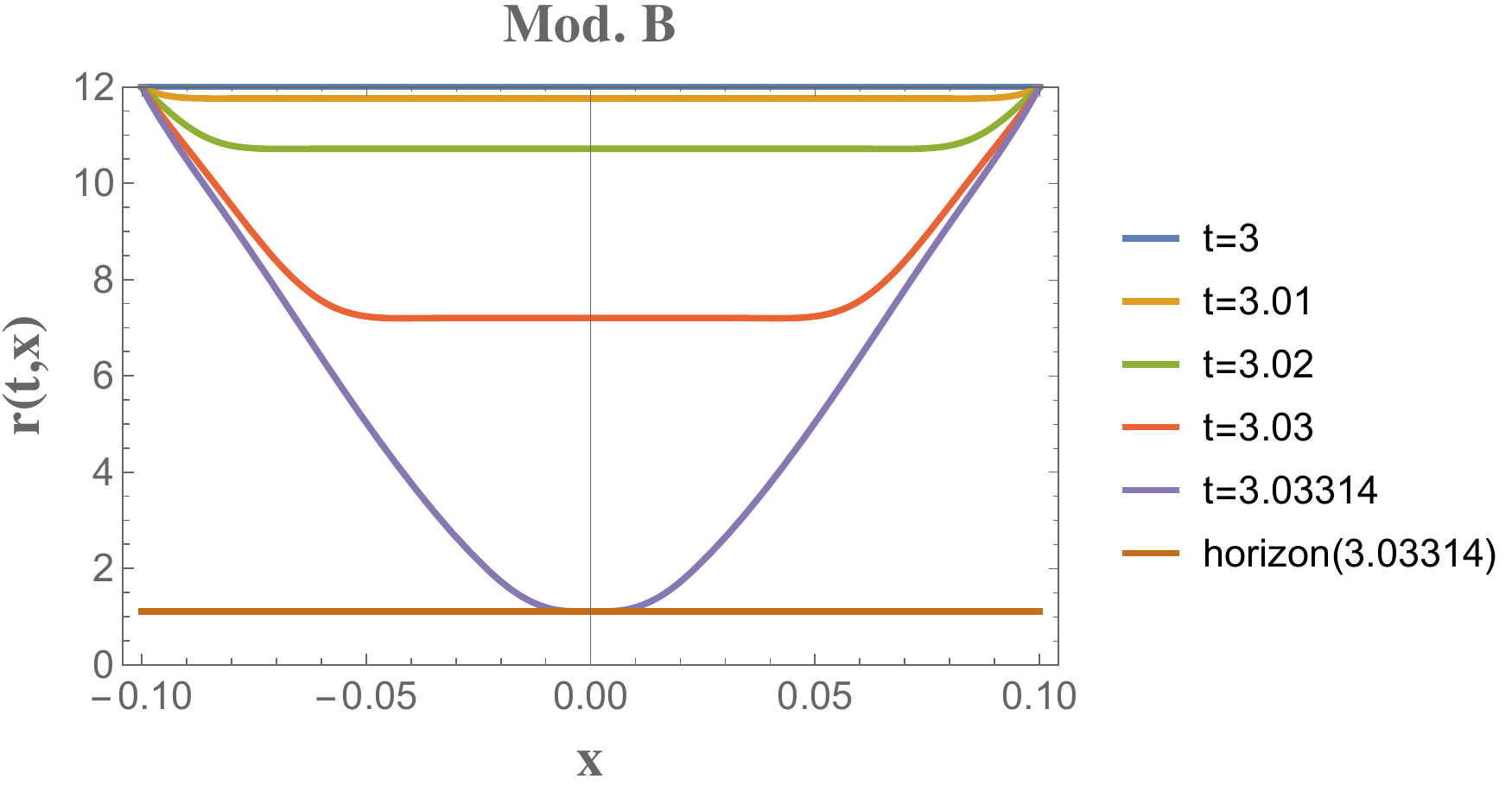}
\includegraphics[width = 0.45\textwidth]{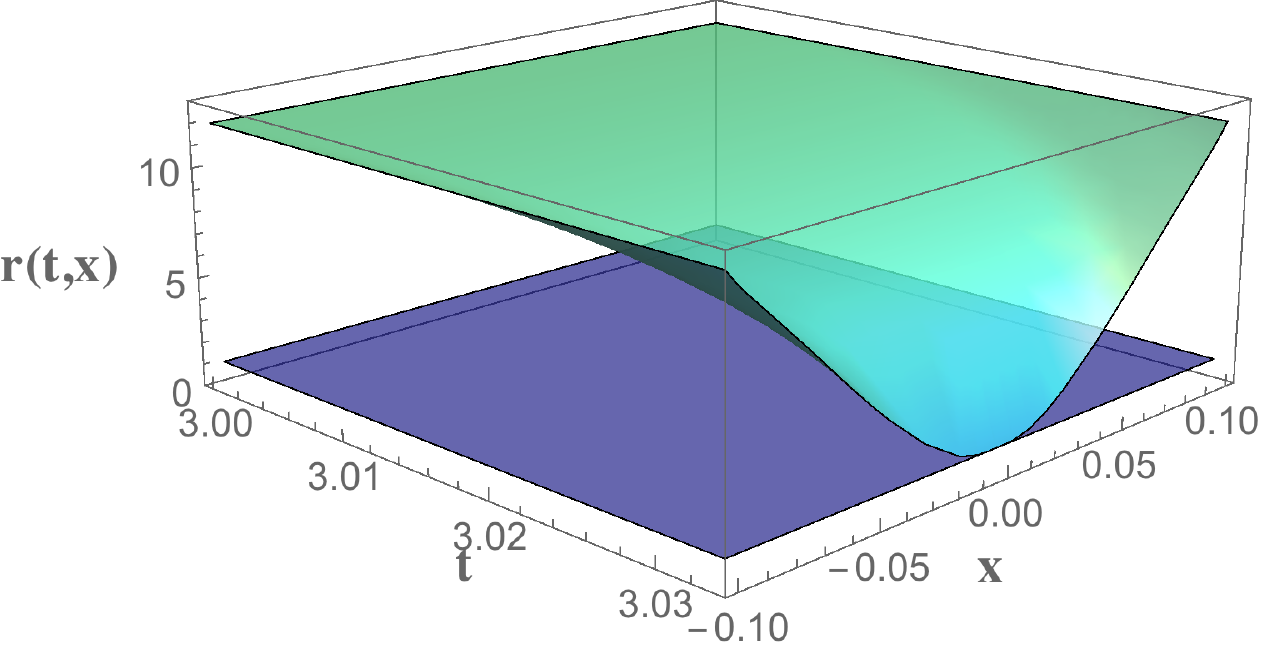}
\caption{\baselineskip 10pt \small 
String profile $r(t, w)$, corresponding to  $\{t_{i}, v, L\} = \{3, -1, 0.1\}$, for the transverse $w = x$ configuration and quench model $\cal B$. Left:   profile as a function of $x$ at different  $t$,  until the horizon is reached. Right:  solution $r(t,x)$  (sea-green surface).   The blue surface corresponds to the horizon. }\label{solutionsA2}
\end{center}
\end{figure}
%
%
%

In Fig.~\ref{solutionsA2} we show the string profile  in the quench model ${\cal B}$, in the case  $w=x$ and separation $L=0.1$. Similar results are found for the  $w=y$ configuration and for model ${\cal A}(2)$. The dissociation time $t_D$ is obtained when   the horizon is reached.

The   dependence of  $t_D$  on the starting time $t_i$  is depicted in Fig.~\ref{columns} for  the two  models.
In each column we draw  the quench profile (upper panel) and $t_D$ for the longitudinal $w=y$ configuration  with  $y_L=10$  (middle panel),
and transverse $w=x$ configuration   with   $L=10$ (bottom panel). In both  models   the dissociation times exhibit abrupt fluctuations during the  quenches, with a tiny delay from the maxima of the pulses. Soon after the end of the quenches, $t_D$ varies smoothly and approaches the same value in both models. During the  quenches, the behavior of $t_D$ is different for the longitudinal and transverse string configurations,  the maxima correspond to minima when switching from one configuration to the other. This is reminiscent of the behavior  of the longitudinal   $p_\parallel$ and transverse $p_\perp$ pressure  \cite{Bellantuono:2015hxa},  maxima in $p_\parallel$  correspond to minima in  $p_\perp$ and viceversa.
 However,    a connection between the two sets of observables justifying the analogous behavior of  pressures and dissociation times is difficult to establish.

Dissociation is faster for $w=x$, with  the string  in a  plane transverse  to the collision axis. 
This is in agreement with the results of several studies suggesting that the orientation of the string and the position of its endpoints are relevant for the dual field interpretation of the quarkonium properties.
For example, in Ref. \cite{Liu:2006nn} it has been found that the direction of the string with respect to the hot wind, namely to the velocity of the fluid produced in the collision, affects the quarkonium screening length, which is minimum for a string oriented perpendicular to the wind and maximum for one parallel to it. Such a pattern is consistent with our results for the dissociation time, since the cases $w=y$ and $w=x$ correspond, respectively, to a quark-antiquark pair parallel and perpendicular to the direction along which the plasma expands. 
Other results with similar interpretation are reported  in \cite{Natsuume:2007vc,Fadafan:2015ynz}.

\begin{figure}[t!]
\begin{center}
\hspace*{1cm}
\includegraphics[width = 0.65\textwidth]{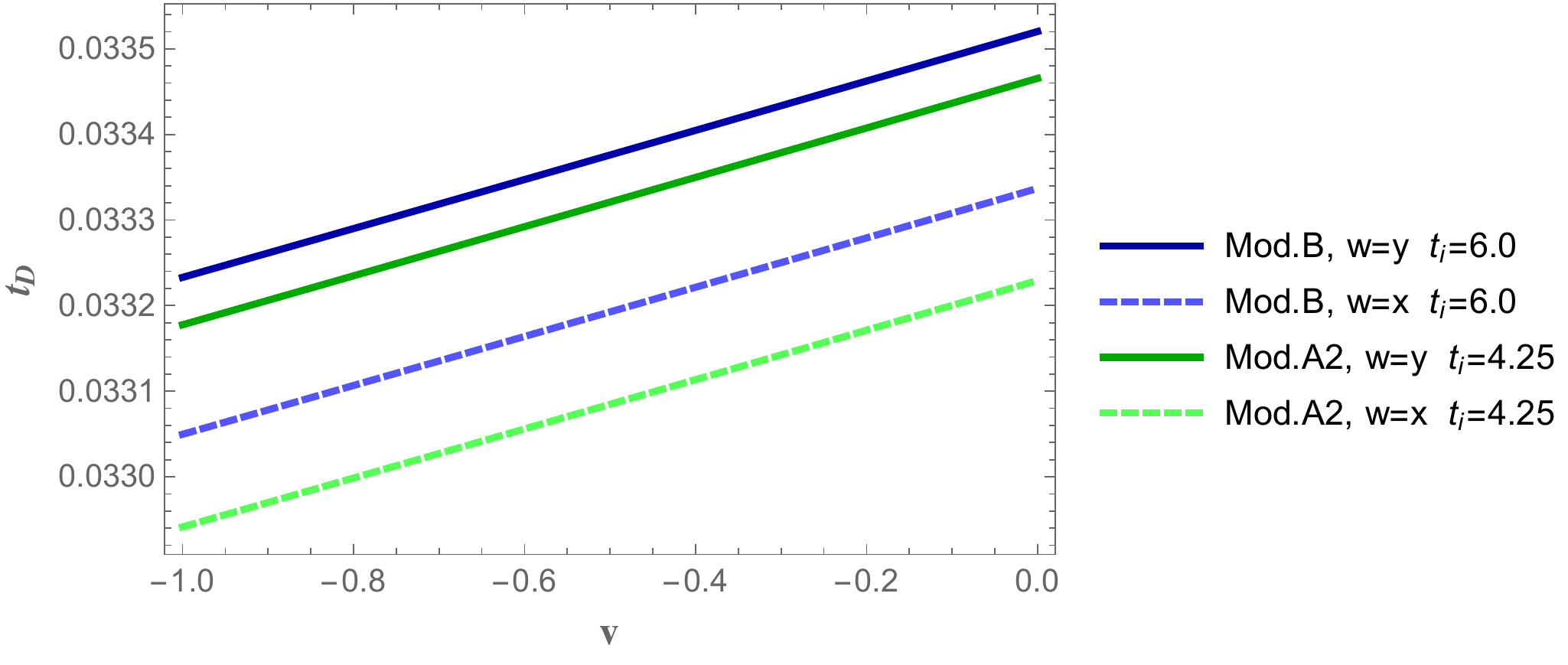}
\caption{\baselineskip 10pt \small Dissociation time $t_D$ versus  $v$  for models ${\cal A}(2)$ and ${\cal B}$, in the transverse $w=x$ configuration with separation   $L=10$ and in the $w=y$ configuration with   $y_L=10$. The initial time is chosen in such a way that the same $\Delta t=1$ has elapsed after the end of the quench for each model.}\label{tDvsV}
\end{center}
\end{figure}

We also observe a mild dependence on the initial velocities:
varying $v$ in the range [-1,0],   a linear dependence of $t_D$  is found for both the quench models and  string configurations, as  shown  in Fig.~\ref{tDvsV} where the initial time  is fixed in such a way that   the same  $\Delta t=1$ has elapsed after the end of the boundary metric distortion. The linear $v$ dependence is expressed in terms of the intercept $a^{M,w}$ and slope $b^{M,w}$,    for the two models $M={\cal B},{\cal A}(2)$ and configurations $w=x,y$:  
$a^{{\cal B},x}=0.0333366$ and  $b^{{\cal B},x}=0.000287307$,
$a^{{\cal B},y}=0.0335199$ and  $b^{{\cal B},y}=0.000287263$,
$a^{{\cal A}(2),x}=0.0332289$ and  $b^{{\cal A}(2),x}=0.000287869$,
$a^{{\cal A}(2),y}=0.0334651$ and  $b^{{\cal A}(2),y}=0.00028788$. The  slopes are compatible with the result of Eq.~(\ref{tAdS}).
%
\begin{figure}[t!]
\begin{center}
\includegraphics[width = 0.45\textwidth]{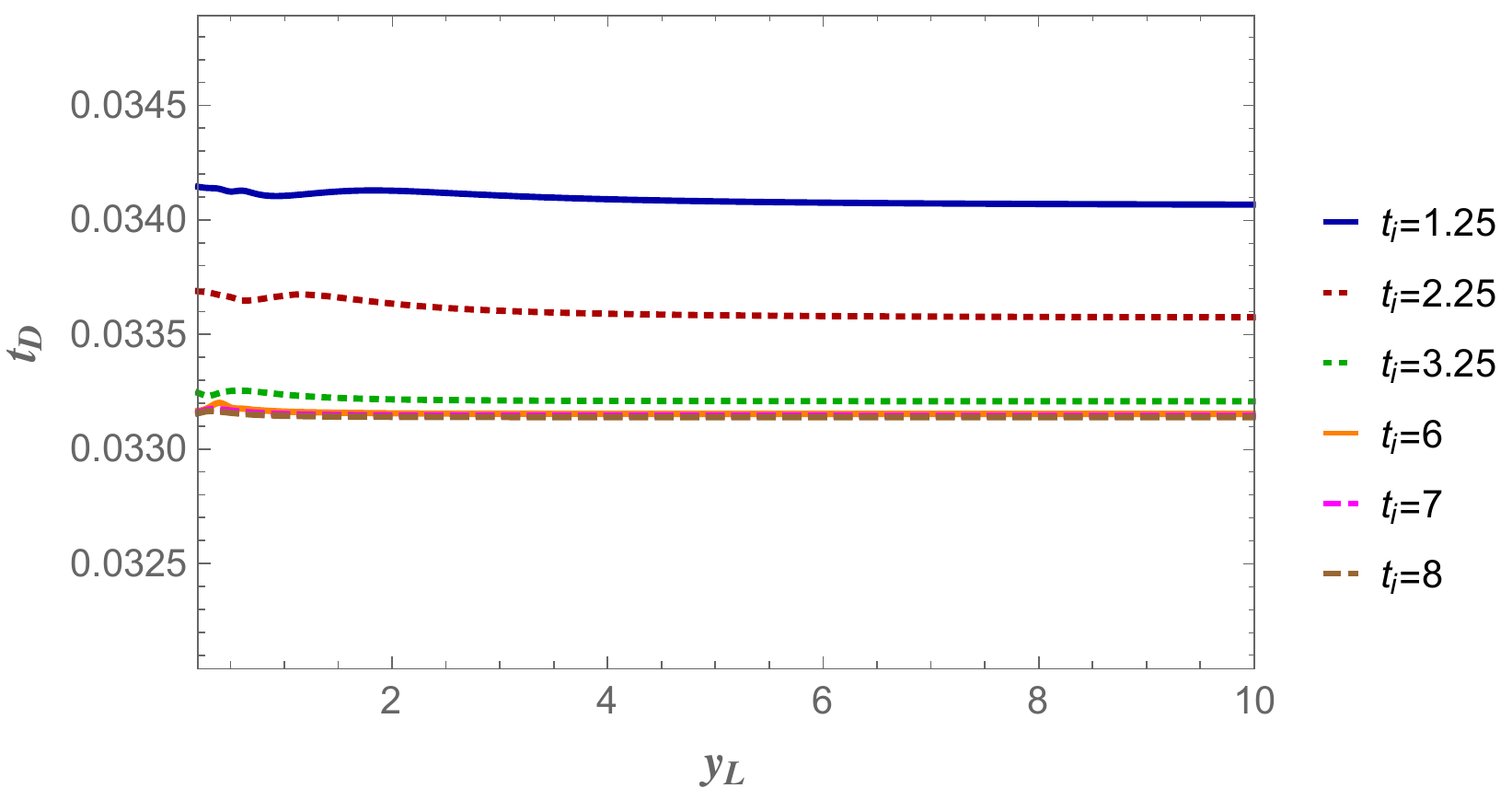}\hspace*{0.6cm}
\includegraphics[width = 0.45\textwidth]{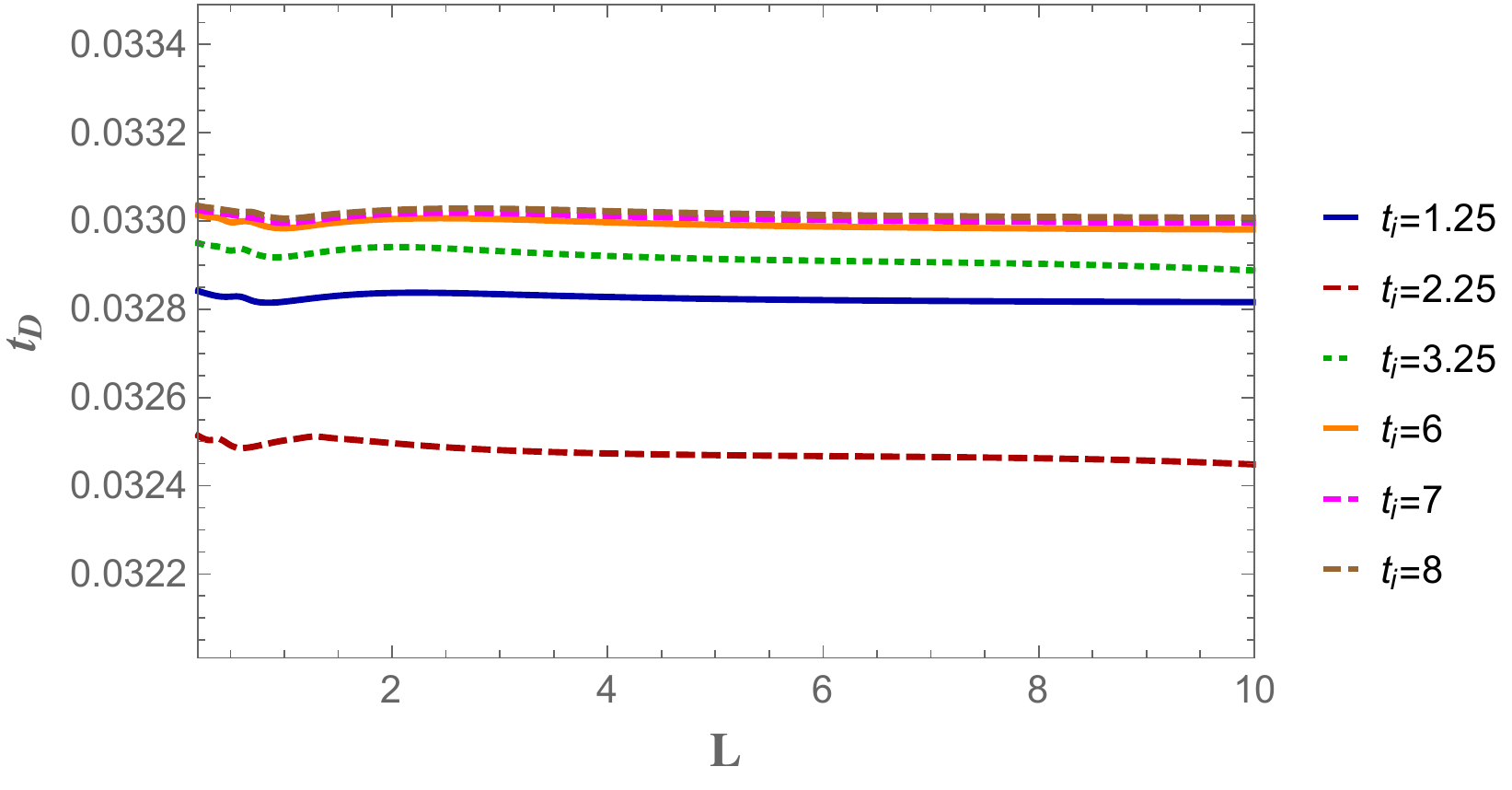}\\
\includegraphics[width = 0.45\textwidth]{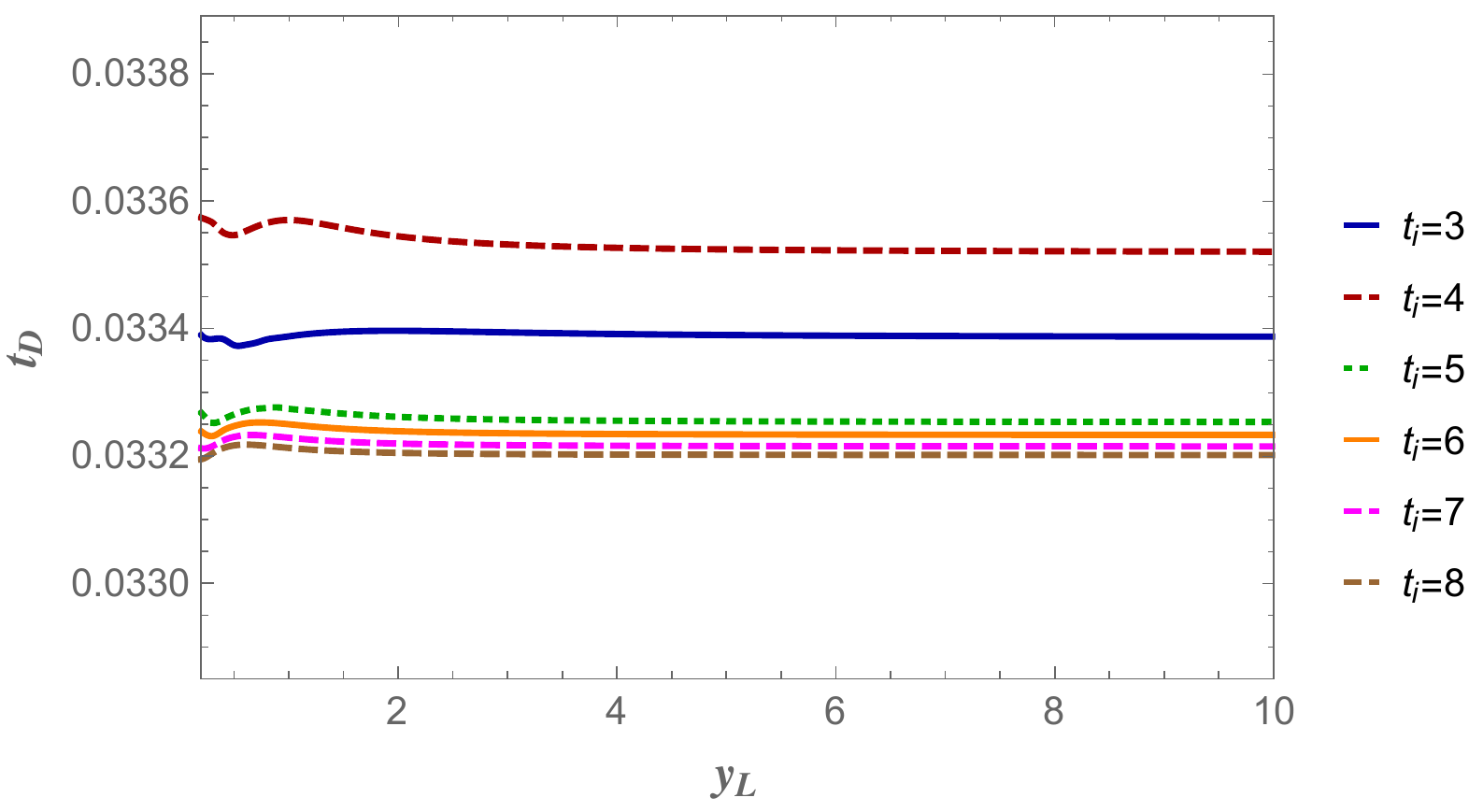}\hspace*{0.6cm}
\includegraphics[width = 0.45\textwidth]{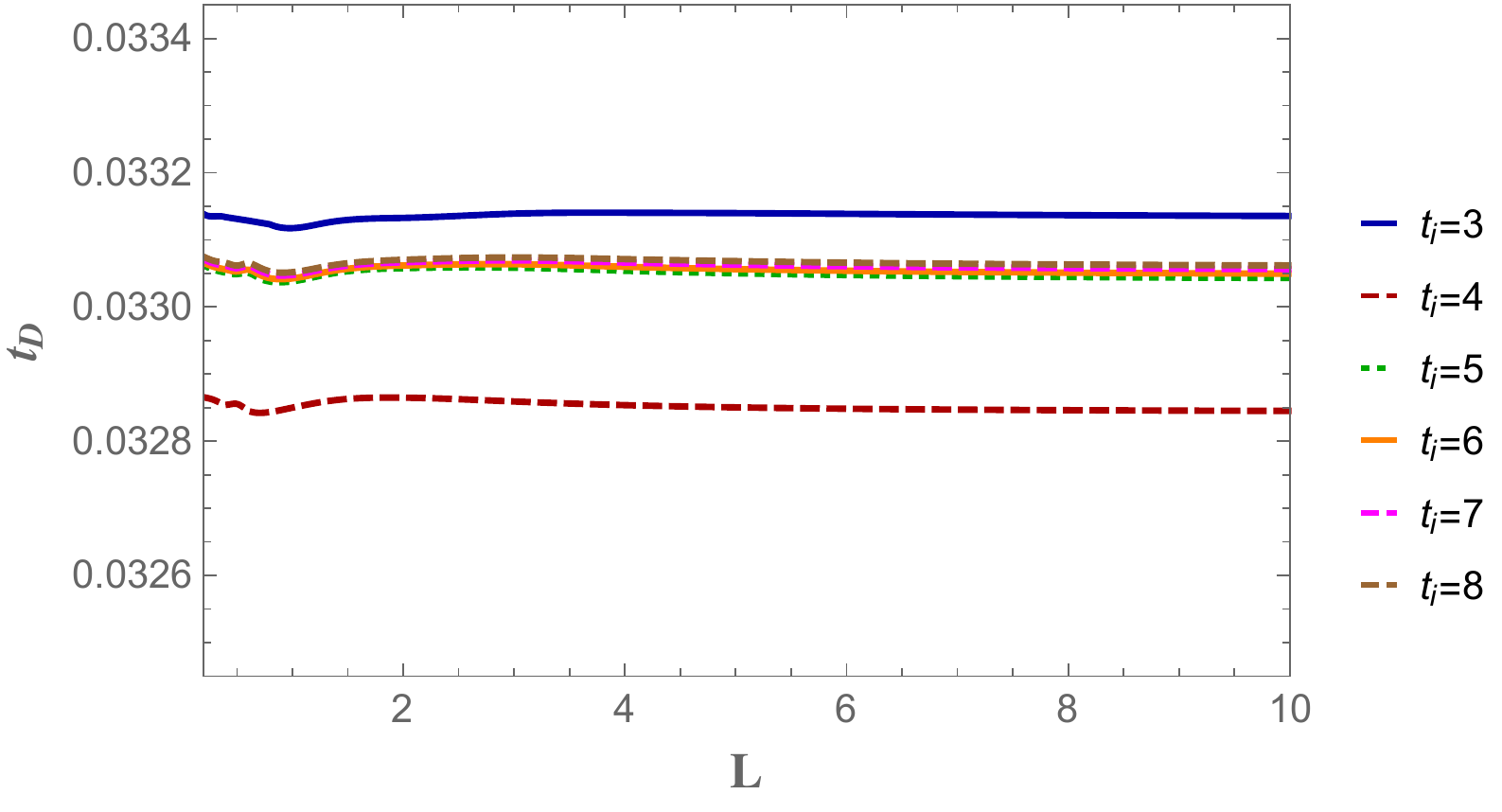}\\
\caption{\baselineskip 10pt \small  Dissociation time  $t_D$   for several values of the initial time, for quench model $\cal A$(2) (upper plots) and  ${\cal B}$ (lower plots),  and    initial velocity $v=-1$. Left:    $t_D$ versus rapidity separation $y_L$ for the  $w=y$ string configurations. Right: $t_D$ versus separation $L$ for the transverse $w=x$ configuration. }\label{tDvsL}
\end{center}
\end{figure}
\begin{figure}[t!]
\begin{center}
\includegraphics[width = 0.43\textwidth]{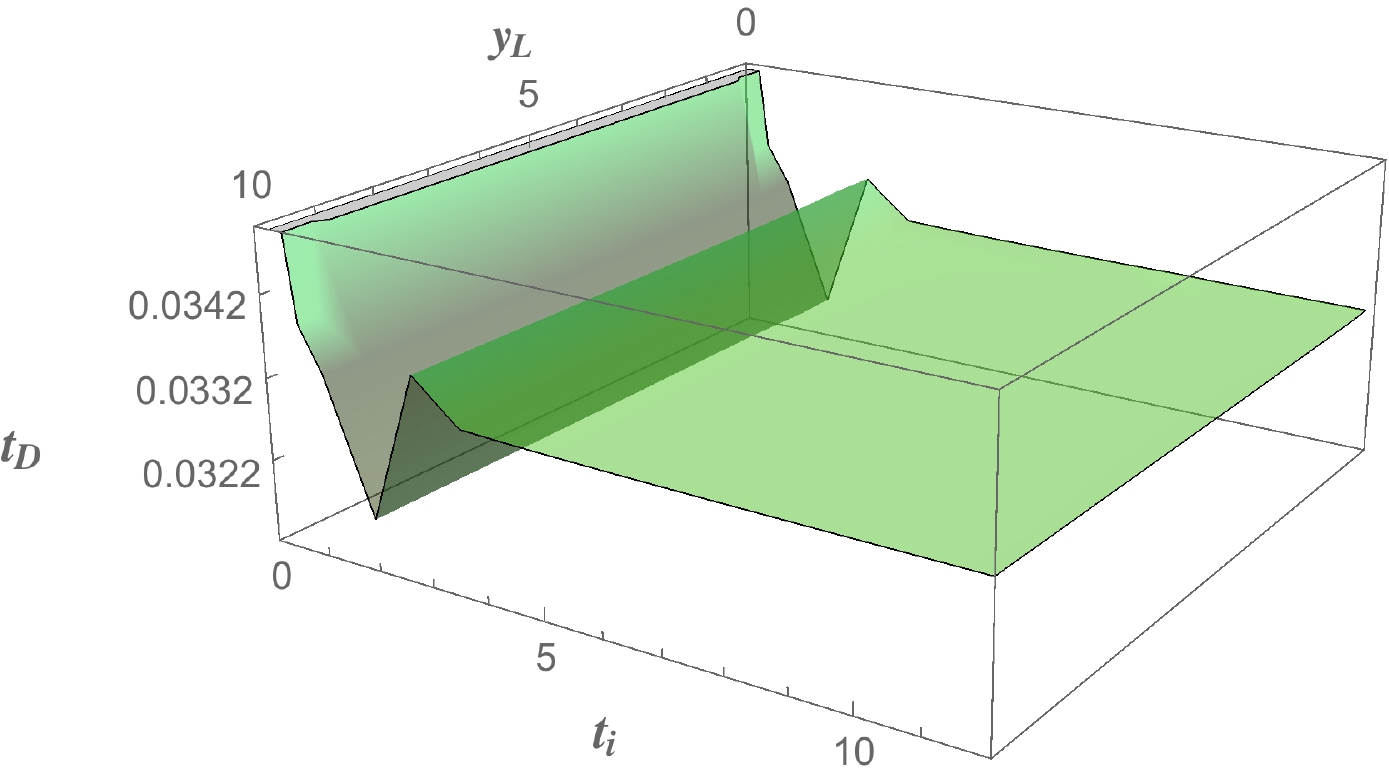} \hspace*{0.6cm}
\includegraphics[width = 0.43\textwidth]{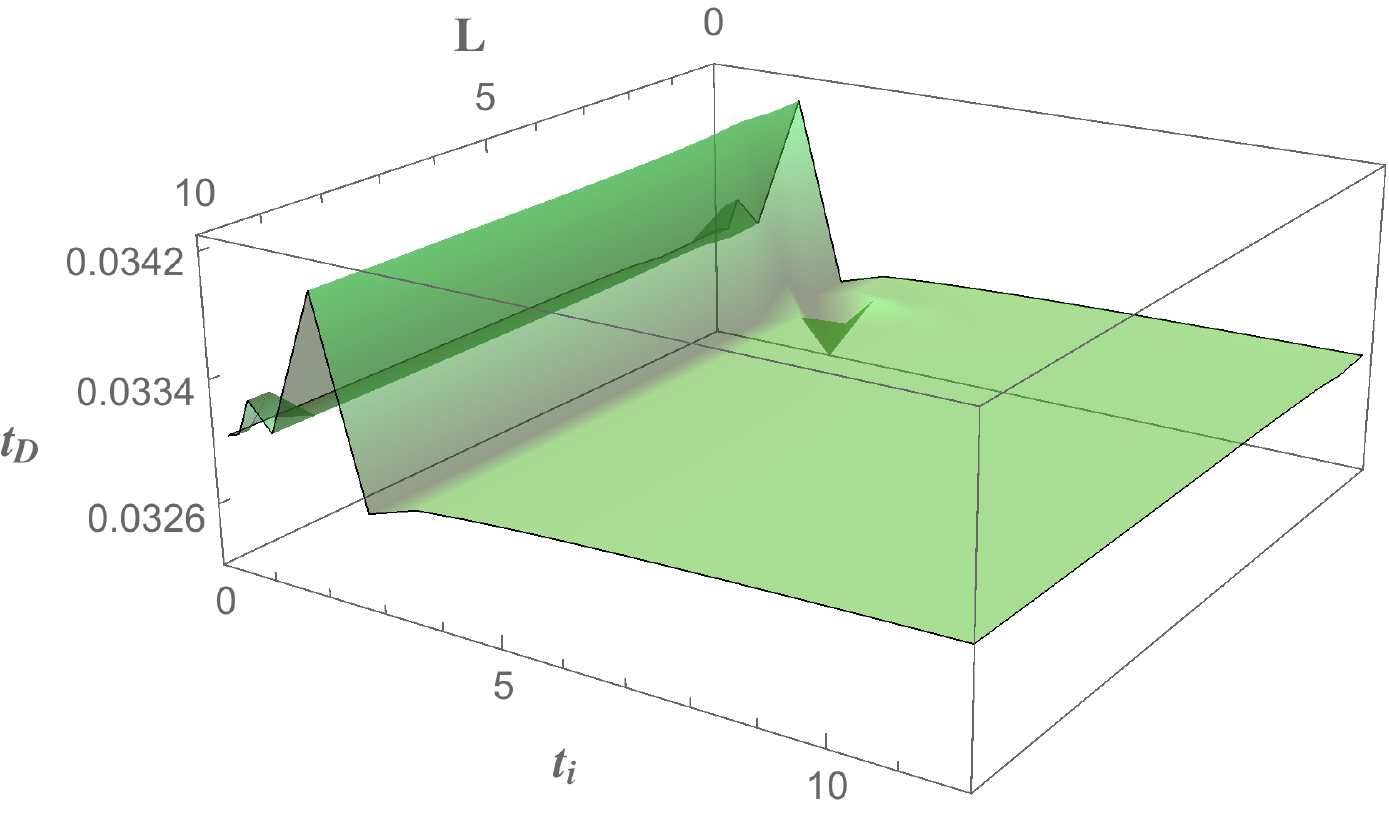}\\
\includegraphics[width = 0.43\textwidth]{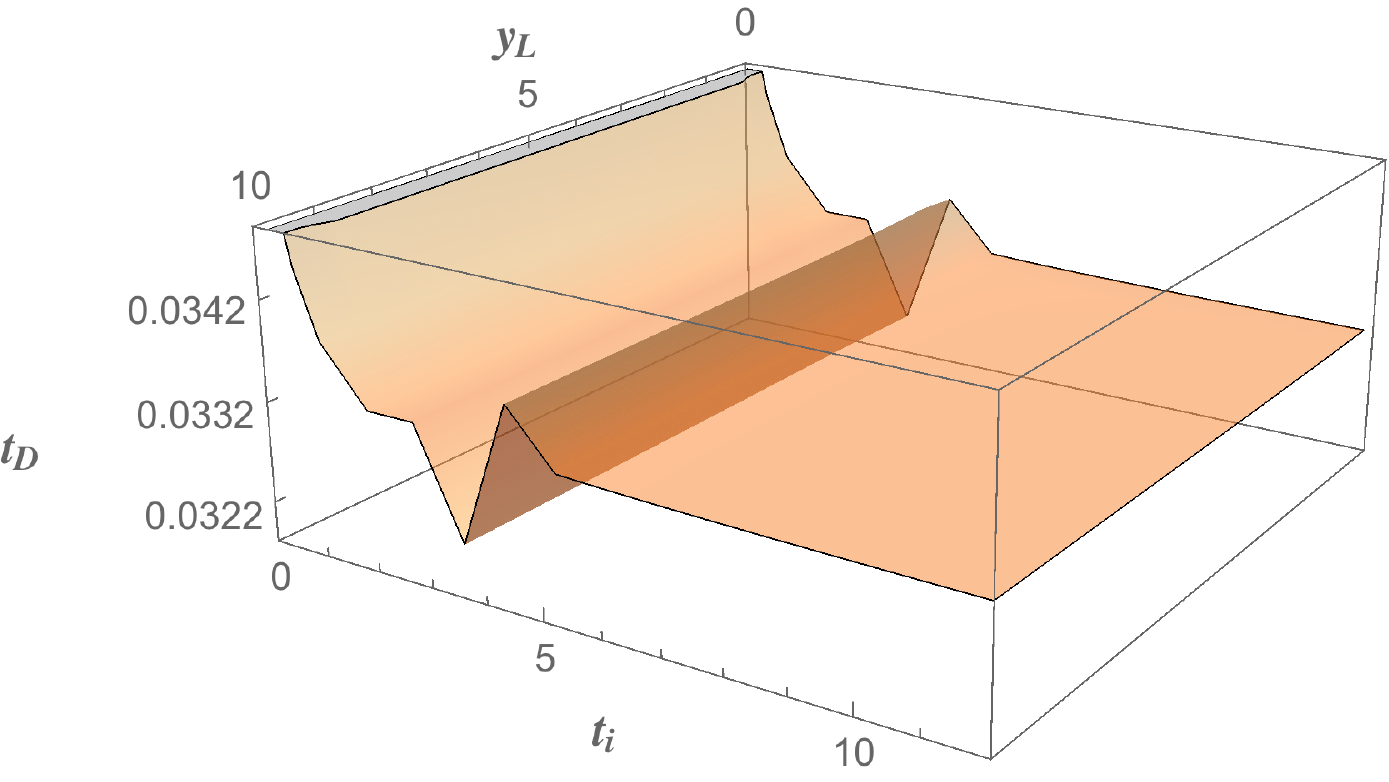}\hspace*{0.6cm}
\includegraphics[width = 0.43\textwidth]{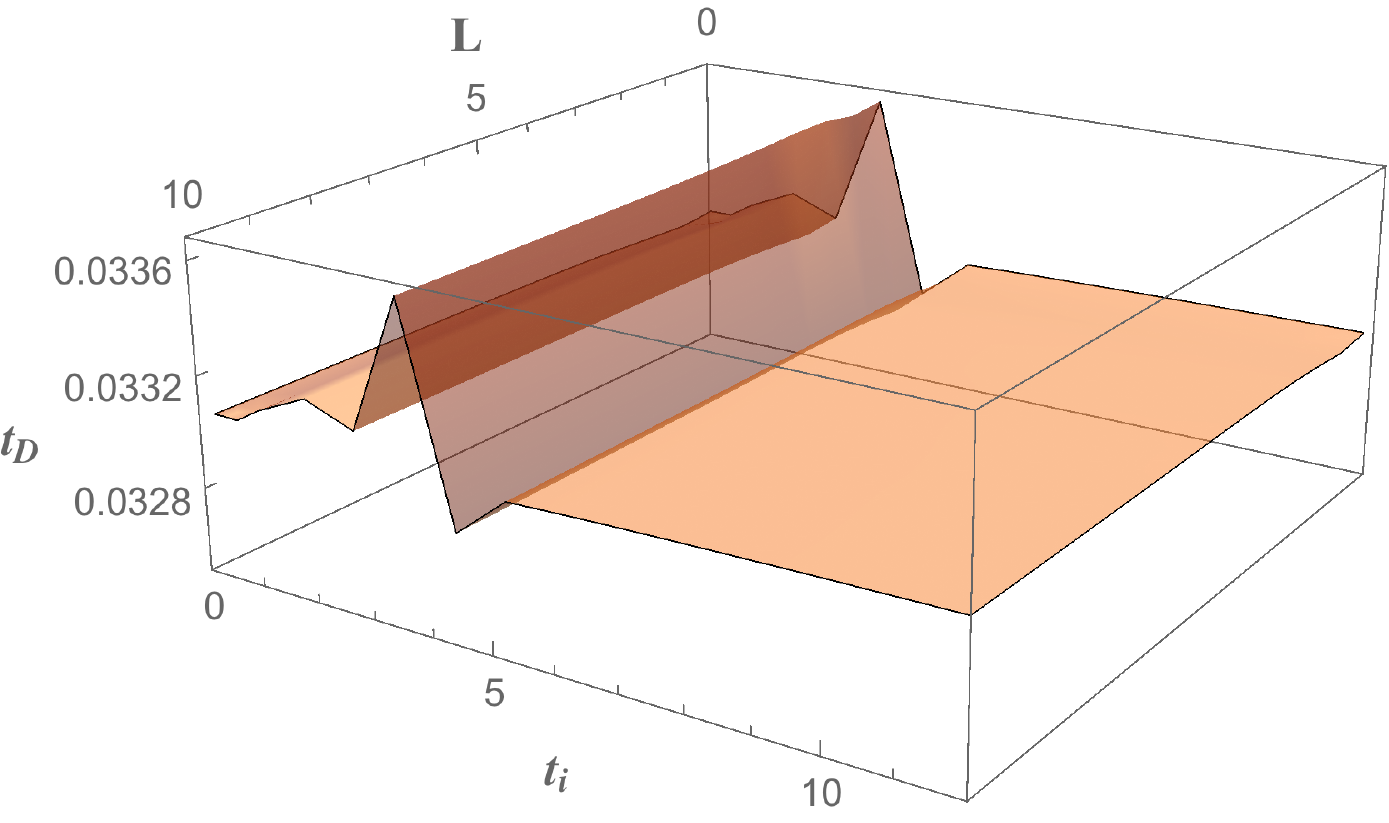}\\
\caption{\small Quarkonium dissociation time $t_D$ for quench model $\cal A$(2) (upper plots) and  ${\cal B}$ (lower plots)  and  initial velocity $v=-1$. Left: $t_D$ vs $t_i$ and rapidity separation $y_L$   for the configuration  $w=y$. Right:   $t_D$ vs $t_i$ and separation $L$ for the transverse $w=x$ configuration.}\label{time3D}
\end{center}
\end{figure}

Varying the  separation between the string  endpoints,  we obtain  the results 
in Figs.~\ref{tDvsL} and \ref{time3D}  for the different models and configurations  and for  selected  initial time. Three values of $t_i$ in Fig.\ref{time3D} correspond to  the start of the first pulse,   the peak and  the end of the last pulse, and three values are taken after the last pulse is switched off.  Our  algorithm is able to determine the dissociation times up to  small values of the separation,  $L=0.1$ and $y_L=0.1$. $t_D$ remains  largely independent of the separation, with some  fluctuations  ascribed in part to numerical effects that can be reduced using a method of changing  variables proposed in \cite{Lin:2006rf}.

Dissociation is  a fast phenomenon. It is strongly affected by the quenches and,  as  in  systems with  nonlinearities,  it does not closely follow the quench profiles. Similarly to some local observables,  namely the energy density, as soon as  the last pulse is switched off 
the behavior in the geometry dual to viscous hydrodynamics  is  recovered, as it can be seen in Fig.~\ref{tDvsHydro}. 
The figure also shows that, although the hydrodynamic model could provide a description of the late time behavior ($t>6$ in model $\cal B$), it is not able to capture the properties of the system  during the rapidly varying regime.
 Setting  the physical units as in \cite{Bellantuono:2015hxa}, assuming a temperature  of the system after the quenches  $T \sim 500$ MeV, the dissociation time is  ${\cal O}(10^{-2})$ fm.

 An  interesting issue to comment on is that, as observed  in \cite{Chesler:2008hg} and \cite{Keegan:2015avk}, in the approach of boundary sourcing
the boundary metric is flat in the  $t=\pm \infty$ limit and  is curved  in correspondence of  the pulses, 
so that is worth asking if the effect of the curved boundary  can be separated from the genuine out-of-equilibrium effect in determining the dissociation times. Although such a separation is difficult to  implement analytically, one can observe that
 for both the considered quench models the time profiles of the boundary curvature invariants 
(Ricci and  Kretschmann scalars) are different from the dissociation time profiles displayed in Fig.\ref{columns}. Moreover, we found  that $t_D$ depends on the string orientation  in correspondence of the pulses.  This difference in time profiles induces us to argue that  the genuine  out-of equilibrium matter effect is captured  in correspondence of the boundary pulses.

\medskip
Before concluding our study of  the real-time quarkonium dissociation dynamics in a far-from-equilibrium plasma, we remark  that in our models  the confinement effects are not accounted for. 
In the holographic framework, a mechanism producing confinement in a strongly-coupled system can be implemented  by a  self-interacting scalar field dual to a relevant deformation of the boundary conformal field theory \cite{Gubser:2008yx}, with a scalar potential   designed to produce expected features of QCD,  a crossover,  a 1st or a 2nd order deconfinement transition \cite{Ishii:2015gia,Janik:2016btb,Janik:2015iry}.  In the critical regions of the QCD phase diagram the role of nonhydrodynamical degrees of freedom seems   more pronounced than in models similar to the one we have considered  \cite{Janik:2015iry}. The description of the real-time quarkonium evolution in such  models, driven out-of-equilibrium, and the comparison with the  results  obtained here  deserve  dedicated analyses.

\medskip
\medskip
\noindent{\bf Acknowledgments} \\
This study has been carried out within the INFN project (Iniziativa Specifica) QFT-HEP.
\bibliographystyle{apsrev4-1}
\bibliography{refs}
\end{document}